\documentclass[aps,prc,reprint
]{revtex4-1}

\usepackage[dvips]{graphicx}
\usepackage{amsmath}
\usepackage{color}      

\begin{document}
\title{Comment on "Structure of the two-neutrino double-β decay matrix elements within perturbation theory" by Du\v{s}an \v{S}tef\'anik, Fedor \v{S}imkovic, Amand Faessler,  arXiv:1506.00835 [nucl-th], Phys.Rev.C91 (2015) 6, 064311}

\author{Vadim Rodin}
\email{vadim.rodin@hotmail.de}
\affiliation{Schmiedewiesen 13, D-72805, Lichtenstein, Germany}
 
\author{O.A. Rumyantsev}
\email{o.rumyantsev@iskfort.ru}
\affiliation{ISK FORT, Prospekt Mira, 102 (1), Moscow, Russia}

\author{M.H. Urin}
\email{urin@theor.mephi.ru}
\affiliation{
National Research Nuclear University “MEPhI” (Moscow Engineering Physics Institute), 115409 Moscow, Russia
}

\begin{abstract}
We comment on a priority claim given by the authors of Phys.Rev.C91 (2015) 6, 064311. 
\end{abstract}

\date{\today}
\maketitle

A false claim of priority has been made by the authors of Ref.~\cite{Stefanik2015}. In the beginning of Sect.~V they have written
"In Ref. [17] the double Fermi and GT sum rules associated with $\Delta Z = 2$ nuclei were introduced.",
where a previous paper~\cite{Stefanik2013} by the same authors has been given the priority (explicit representations for the sum rules are given in 
Eqs.(27),(28) of~\cite{Stefanik2015}).

In fact, the sum rules were first introduced in~\cite{RumUr98} 15 years prior to the publication~\cite{Stefanik2013}, see Eq.(5) of~\cite{RumUr98} (the notation used in~\cite{RumUr98} is somewhat different from that of~\cite{Stefanik2013,Stefanik2015}). 
A model-independent, identity transformation of the amplitude $M^{2\nu}$ of $2\nu\beta\beta$ decay, introduced for the first time in ~\cite{RumUr98}, allows one to partition $M^{2\nu}$ into two terms that are sensitive to different parts of a nuclear Hamiltonian. One of these terms is proportional to the sum rule in question.
As the authors of~\cite{RumUr98} only considered a mean field contribution to the sum rules, they concluded that the sum rules must vanish, that is not correct for a general nuclear Hamiltonian.
This shortcoming of~\cite{RumUr98} was corrected in follow-up papers~\cite{Rodin2005,Rodin2011}, where the dominating contribution of the particle-particle sector of a nuclear Hamiltonian to the sum rules was emphasized and evaluated in the quasiboson approximation. A simple separable particle-particle interaction was employed in~\cite{Rodin2005}, whereas the case of a general particle-particle interaction was treated in~\cite{Rodin2011}.

Note, that the authors of~\cite{Stefanik2013,Stefanik2015} must have been aware of publications ~\cite{Rodin2005,Rodin2011}, since first,~\cite{Rodin2005,Rodin2011} were cited in \cite{Stefanik2013}, see entries [26,27] in the list of references, and second, one of the authors of~\cite{Stefanik2013,Stefanik2015} was also a co-author of~\cite{Rodin2005,Rodin2011}.



\begin{thebibliography}{10}

\bibitem{Stefanik2015} 
  D.~\v{S}tef\'anik, F.~Simkovic and A.~Faessler,
  Phys.\ Rev.\ C {\bf 91}, no. 6, 064311 (2015)

\bibitem{Stefanik2013} Du\v{s}an \v{S}tef\'anik
  D.~\v{S}tef\'anik, F.~\v Simkovic, K.~Muto and A.~Faessler,
  Phys.\ Rev.\ C {\bf 88}, no. 2, 025503 (2013)

\bibitem{RumUr98} O.~A.~Rumyantsev and M.~H.~Urin,
  Phys.\ Lett.\ B {\bf 443}, 51 (1998).

\bibitem{Rodin2005} V.~A.~Rodin, M.~H.~Urin and A.~Faessler,
  Nucl.\ Phys.\ A {\bf 747}, 295 (2005).

\bibitem{Rodin2011} V.~Rodin and A.~Faessler,
  Phys.\ Rev.\ C {\bf 84}, 014322 (2011)

\end{thebibliography}
\end{document}